# New Ways to Synthesize Lead Sulfide Nanosheets –

# Substituted Alkanes Direct the Growth of 2D Nanostructures

Thomas Bielewicz, Eugen Klein, and Christian Klinke*

*Institute of Physical Chemistry, University of Hamburg,*

*Grindelallee 117, 20146 Hamburg, Germany*

**Abstract**

*Two-dimensional colloidal nanosheets represent very attractive optoelectronic materials. They combine good lateral conductivity with solution-processability and geometry-tunable electronic properties. In case of PbS nanosheets, so far the synthesis was driven by the addition of chloroalkanes as coligands. Here, we demonstrate how to synthesize two-dimensional lead sulfide nanostructures using other halogen alkanes and primary amines. Further, we show that at a reaction temperature of 170 °C a coligand is not even necessary, and the only ligand, oleic acid, controls the anisotropic growth of the two-dimensional structures and using thiourea as sulfide source, nanosheets with lateral dimensions of over 10 μm are possible.*

* Corresponding author: klinke@chemie.uni-hamburg.de

## 1. Introduction

Nowadays, the colloidal synthesis is an established method to produce solution-based, monodisperse nanostructures with tunable size, shape, and composition. The properties of such materials are strongly influenced by the nature of the defining organic ligands.[1] $Cu_2S$ platelets, for example, can be controlled to yield large triangular or hexagonal shapes by using different halides and thiols as ligands.[2] For other materials, like CdSe or ZnO nanosheets,[3,4] long-chained alkanes with a polar head group are the crucial compounds to form two-dimensional nanocrystals. Two-dimensional colloidal nanosheets represent very attractive optoelectronic materials.[5,6,7] They combine good lateral conductivity with solutions-processability and geometry-tunable electronic properties. In case of PbS nanosheets, the synthesis was shown to be driven by the addition of chlorinated alkanes as coligands.[8,9,10,11] Recently, we could show that different chloroalkanes can lead to similar lead sulfide nanosheets and only the reaction temperature or the amount of the chloroalkane plays a role for the formation of two-dimensional nanocrystals.[10] An interplay between oleic acid/oleate and the chloroalkane with the crystal facets of lead sulfide was identified as the main reason for small lead sulfide nanocrystals to merge via oriented attachment rather than to form individual spherical or cubic nanoparticles.[11] The reason why the growth to laterally large sheets with thicknesses smaller than the Bohr radius of PbS should be exclusive to the use of chloroalkanes is not investigated until now.

We use different halogen alkanes as well as amines with two different chain lengths to investigate the effect of those weakly binding coligands on the shape of lead sulfide. Further, we elucidate why some substituted alkanes lead to two-dimensional structures while others yield spherical or star-shaped nanoparticles. To systematically study the effect of the ligands, we used fluoro-, chloro-, and bromoalkanes as well as amines with carbon chain-length of C7 and C14. The reaction temperature in all experiments was 170 °C. Two different sulfide sources were used in this study, the more reactive thioacetamide (TAA) and the less reactive thiourea (TU).

## 2. Experimental method

*Synthesis*

All chemicals have been used as received. The chemicals used were: lead(II) acetate trihydrate (Aldrich, 99.999 %), thioacetamide (Sigma-Aldrich, >= 99.0 %), thiourea (Aldrich, >= 99.0 %), diphenyl ether (Aldrich, 99 %+), dimethyl formamide (Sigma-Aldrich, 99.8% anhydrous), oleic acid (Aldrich, 90 %), 1-chlorotetradecane (Aldrich, 98 %), 1-chloroheptane (Aldrich, 99 %), 1-fluorotetradecane (Lancaster, 98 %), 1-fluoroheptane (Aldrich, 98 %), 1-bromotetradecane (Aldrich, 97 %), 1-bromoheptan (Aldrich, 99 %), 1-iododecane (Aldrich, 98 %), tetradecylamine (Aldrich, 95 %) and heptylamine (Aldrich, 99 %).

A 50 mL four neck flask with a thermocouple, a thermometer, a condenser and septum was used for all experiments. 860 mg of lead acetate trihydrate (2.3 mmol) was dissolved in 10 mL diphenyl ether and 3.5 mL of oleic acid (10 mmol). The solution was heated to 75 °C and a vacuum was applied for two hours at 0.3 mbar. Then, under a nitrogen atmosphere, the solution was heated to 100 °C where 0.7 mmol or 2.9 mmol of the coligand (alkylamine or fluoroalkane or chloroalkane or bromoalkane) was added rapidly. After that, the temperature was increased to the reaction temperature of 170°. The molar amount of the coligand was 0.7 mmol and 2.9 mmol, resp. In the reaction with tetradecylamine it was first dissolved in 2 mL of diphenyl ether and heated to 60 °C before it was injected into the solution. To start the reaction, 0.2 mL of a 0.04 g TAA (0.5 mmol) in 6.5 mL DMF was added at 170 °C. The solution was stirred at 700 rpm for 5 min before cooling down to room temperature. To separate the product from the solvent and byproducts, the reaction solution was centrifuged at 4000 rpm for 3 min and the precipitant washed two times with toluene. The product was stored suspended in toluene.

*Methods*

*TEM*

The TEM samples were prepared by diluting the nanostructure suspensions with toluene and then drop casting 10 µL of the suspension on a copper TEM grid coated with a carbon film. TEM images were taken using a JEOL-1011 with a thermal emitter operated at an acceleration voltage of 100 kV.

*XRD*

X-ray diffraction measurements were performed on a Philips X'Pert System with Bragg-Brentano geometry and a copper anode with an X-ray wavelength of 0.154 nm. The samples were measured by drop-casting the suspended nanosheets on a <911> or <711> grown silicon substrate.

*Simulations*

For the interpretation of the results we performed simulations in the frame of the density functional theory (DFT) to evaluate the ligand absorption on the different facets of PbS by employing the ORCA software.[20] The calculations were performed with the LANL ECP[21]/ def2-TZVP/J[22] basis set for Pb and the def2-TZVP/J one for the lighter elements. Furthermore, the LDA exchange functional and the correlation functional VWN-5[23] are used as DFT framework. To keep the simulation time at a feasible level, we simulated shorter versions of the ligand molecules with the functional group and a three carbon aliphatic chain (e.g. instead of oleic acid we simulated propanoic acid). This does not change the qualitative aspects of the results. The adsorption energies were calculated by comparing the sum of the separate energies of the PbS crystal and the ligand molecule with the total energy of the complete system. During simulation under aperiodic boundary conditions, the PbS crystal was fixed to the experimental lattice constant and the geometry of a pristine facet while the ligand molecule was free to relax.

## 3. Results and discussion

### *C7 ligands*

The molar amounts of the coligands were chosen to be less (0.7 mmol) respectively more (2.9 mmol) than the molar amount of the lead source (2.3 mmol). The negatively polarized head groups of the coligands interact with the positively charged lead ions on the surface of the crystals. By using the seven carbon long coligands at a temperature of 170 °C and TAA as the sulfide source the fluoro, chloro, and amine alkanes yielded two-dimensional nanostructures. 1-bromoheptan (BH) was the only alkane which yielded spherical nanoparticles only, independent of the molar amount used. The XRD pattern of latter shows all peaks of a *galena* crystal but with non-bulk intensity distribution (Figure 1). After washing the solution thoroughly the major part of the ligands is removed and the nanocrystals arrange randomly causing *galena* typical XRD intensities. Applying this procedure to the obtained nanosheets, which exhibit only peaks for the {100} planes in the XRD patterns, no change in the diffractogram is visible (Figure S1). This means that in the case of nanosheets the texture effect is due to the shape of the inorganic part and cannot be changed by removing the ligands. Thus, XRD is a convenient and fast method to analyze the shape of the product of a lead sulfide synthesis. TEM images allow a further characterization of the difference in influence of the coligands on the synthesis of two-dimensional nanosheets; of interest are the lateral dimensions, smoothness and stacking (Figure 2). 1-chloroheptane (CH), as a coligand, produced square-like sheets with lateral dimensions of over 2 µm for both molar amounts (0.7 mmol and 2.9 mmol). For the fluoro alkane and the amine alkane changes in the shape of the nanosheets are visible for the different molar amounts: With 0.7 mmol of 1-fluoroheptan (FH) the sheets are around 400 x 200 nm$^2$. With 2.9 mmol this changes drastically to over 2 µm for the longer and to over 1 µm for the shorter dimension. The tendency for quadratic and laterally shorter nanosheets at a molar amount of 0.7 mmol can be compared to the CH coligand where also quadratic sheets are obtained with 0.7 mmol. Laterally larger nanosheets are obtained with 2.9 mmol in both cases, with CH or FH as the coligand. Both coligands have the same effect on the lateral dimensions by increasing their molar amount. Nanosheets produced with FH are generally smaller in comparison with sheets produced with CH. Less attractive forces to the lead sulfide crystal (Table S1) lead to a reaction solution where smaller nuclei can precipitate. More nanosheets can be synthesized at the beginning of the reaction but in turn, they cannot grow laterally large. This effect could be similar to the formation of spherical nanoparticles where higher amounts of ligands produce larger spheres and in the case of nanosheets stronger ligands produce laterally larger structures. The sheets obtained with heptylamine (HA) at a molar amount of 0.7 mmol are comparable to the sheets obtained with FH at the same concentrations but with 2.9 mmol of HA the shape changes drastically, too. Stacked, stripe-like structures are the dominant product and XRD shows a small peak for the (311) facet at *2θ = 50°* (Figure 3). Other authors report on a hydrothermal method to produce PbS nanosheets using amines,[12] but those structures possess a thickness of over 50 nm, while those reported here are thinner than 10 nm, calculated using the Scherrer equation.[13]

Lead sulfide nanosheets are formed through oriented attachment where ultra-small nanoparticles merge via the {110} facets of the crystal to minimize the surface energy of the whole nanocrystal. Since oleate, which forms monolayers in the crystalline phase,[14] is also present in the reaction the oriented attachment takes place in two dimensions only because the monolayers of oleate are established on the {100} facets and prohibit a growth or attachment in this direction. The coligands (in this case FA, CH, and HA) are necessary to destabilize the {110} facets and to remove the oleate from these facets, such that oriented attachment can take place; otherwise the oleate would block the facet completely. Oleate as well as the coligands have the highest binding energy with the {110} facets, the strongest binding energy resulting for the oleate (Table S1). The calculated energies alone cannot explain the different shapes obtained in the experiments. For example, BH produces spherical nanoparticles only, independent of the molar amount, but HA, which has a similar binding energy on the {110} facet, can produce two-dimensional sheets. Here, the chemical stability of the coligand has to play a role, too. This is shown in experiments where small amounts (2.9 mmol) of iododecane were enough to yield $PbI_2$ instead of the desired PbS in the presence of the sulfide source (Figure S2-A). We were also able to force the reaction to yield $PbBr_2$, instead of PbS (Figure S2-B). For that, the bromoalkane had to be used as the solvent (10 mL). In contrast, it was not possible to obtain $PbCl_2$ using a chloroalkane, even if used as the solvent (10 mL). With iodide and bromide being much better leaving groups than chloride and fluoride or amine the interaction between the coligand and the lead oleate complex has also to be taken into account. The reaction to form lead bromide or lead iodide is a competing side reaction for the formation of lead sulfide and the attraction to the lead cation changes the interaction of the coligand with the crystal surface significantly and the shape of the product, too. HA is the only coligand with a binding energy for sulfur-rich facets which is higher than the binding energy for lead-rich facets (Table S2). This means it prefers the sulfide spots on the crystal and it does not compete as much with the oleate for the lead. It can still produce two-dimensional nanosheets at lower molar amounts while at higher molar amounts of HA the shape changes to stacked stripes which means that the coligand is more present on the crystal to stabilize the facets over which oriented attachment takes place and complements oleic acid or oleate (due to the higher amounts in the solution) rather than competing with it. Therefore the product becomes stripe-like at lower coligand amounts with HA because it is a stronger coligand. This is supported by our previous findings and experiments done in this study when more chloroalkane is introduced the obtained nanosheets look more stripe-like until they change shape to stripes eventually. Higher amounts of chloroalkane leads to more stabilized crystal facets and oriented attachment gradually takes place in only one lateral dimension.[10] With HA this transition between nanosheets and stripes takes place at lower concentrations of HA and the product changes already from nanosheets at 0.7 mmol of HA to stripe-like agglomerates at 2.9 mmol.

### *C14 ligands*

We could already show that the long-chained chlorotetradecane (CTD) could be used to yield two-dimensional lead sulfide nanosheets.[10] We performed the experiments also with the other alkane coligands with lengths comparable to the length of the main ligand, oleic acid. All results are shown in Figure 4. 1-bromotetradecane (BTD) yields spherical nanoparticles only but with all other coligands we were able to produce nanosheets, too. Comparing the shorter coligand HA with the longer tetradecylamine (TDA) shows that it is possible to obtain nanosheets at a molar amount of 0.7 mmol with both coligands. The sheets with TDA are highly stacked but XRD still shows only peaks for the {100} planes. Using 2.9 mmol of HA deteriorates the shape of the nanosheets and more stacked stripe-like structures are obtained while with TDA the structures resemble more platelets where no ordered structures can be seen (Figure 4). The XRD of the product obtained with 2.9 mmol TDA shows all peaks of *galena* and TEM images show small agglomerated, amorphous platelets. The XRD of the product with 0.7 mmol TDA on the other hand shows dominant {100} peaks but with a narrower (200) peak (Figure 5). Using the Scherrer equation to calculate the sheet thickness, by fitting the (200) peak, also shows that the TDA sheets are much thicker (around 48 nm) than the HA (10 nm) ones. Using fluorinated alkanes, 1-fluorotetradecane (FTD) shows also comparable results whereat a molar amount of 0.7 mmol sheets are obtained which are similar to the results with the shorter coligand FH. For the 2.9 mmol amounts the results differ. In the case of the FH it was possible to obtain laterally larger sheets while with FTD no sheets are formed but agglomerated stripe-like structures. The XRD shows all *galena* peaks with 2.9 mmol of FTD used in the reaction while at a molar amount of 0.7 mmol the {100} planes are the only peaks and nanosheets are produced. This is very similar to our previous findings where more coligand leads to stripe-like strucutres when all other parameters remain the same. [10] CTD is the only long chained ligand which yields sheets with both molar amounts. With 0.7 mmol the nanosheets are square-like and resemble very much the nanosheets obtained with CH. With 2.9 mmol of CTD the sheets have a prolonged lateral dimension and no visible spherical nanoparticles are in the product in contrast to the 0.7 mmol reaction. The XRD shows no crystal planes beside the {100} in contrast to the XRD for the 0.7 mmol reaction where the (220) plane is slightly visible and the TEM images show agglomerates of spherical nanoparticles.

In comparison, the C14 coligands behave similar to the shorter C7 coligands, except for the aliphatic amine and fluoro compound. With a longer carbon chain interactions between it and the PbS crystal or coligand–coligand interactions get more complicated and harder to predict. For example, the longer chains keep the nanoparticles sterically better separated.[15] For oriented attachment to take place the small nanocrystals must make contact with each other and align their crystal facets perfectly. With longer carbon chains this is more difficult to achieve because the surface is passivated and a closer approach is hindered. CTD is the only long chained coligand enabling nanosheet growth with both chosen molar amounts. Using TDA and FTD in higher molar amounts produce agglomerated stripe-like structures or amorphous platelets while BTD produces spherical nanoparticles only. The molar coligand range for nanosheets is slightly shifted for using TDA and FTD in comparison to the shorter HA and FH where nanosheets are formed at both molar amounts.

*Influence of the sulfide source*

An important factor for the formation of lead sulfide nanosheets is the amount of sulfide introduced to the reaction. A 140/1 ratio between lead acetate and the sulfur source is used to direct the product to oriented attachment and eventually to lead sulfide nanosheets. Using TAA at 170 °C has the advantage that TAA decomposed quasi instantly and completely at the beginning of the reaction – and at later stages of the reaction, especially the growth phase, very little sulfide monomer is present. This leads to oriented attachment as the main driving force for ultra-small nanoparticles to reduce their surface energy, since not enough sulfide monomer is present during the growth step, which would lead to growth of spherical nanoparticles. By introduction of a more stable sulfide source, TU,[16] it was not possible to obtain nanosheets as the only product. Spherical nanoparticles were present in all reactions and XRD showed at least one other crystal plane beside the {100}. TU is a sulfide source which decomposes gradually at 170 °C which leads to an excess of sulfide monomer during the growth step. These conditions favor spherical nanoparticle growth to minimize the energy of the crystal surfaces. It has to be noted that the sheets obtained with TU (especially with FH, CH and CTD) were laterally much larger than the ones obtained with TAA as the sulfide source (Figure 6). The reason for this could be similar to spherical nanoparticle growth where a smaller concentration of nanocrystals at the beginning of the reaction leads to larger diameters for the product. The shorter side of the large sheets is more than 6 μm and the longer can reach up to 10 μm. For other reactions sheets which look like grown together stripes were the main product because the coligands can stabilize the small nanoparticles at the beginning of the reaction much better when the sulfide concentration is not as high at the beginning as it is with TAA. Though, it is still high enough to ensure that oriented attachment takes place as well, beside spherical particle growth. Interestingly, the only product achievable with FTD is lead sulfide stars. In XRD (Figure 7), these six-armed stars show more pronounced (111) peaks. This might be a texture effect as well, like with sheets, since the stars have a favorite orientation on the substrate – they are sitting with three of their arms on the substrate. We can assume that the arms are growing from {100} planes,[16,17] the planes parallel to the substrate would then be the {111} planes and have a higher chance to be detected by the XRD. After washing the product several times to remove excess ligand from the surface of the crystals the XRD intensities did not change (Figure S3), which supports the notion that the arms grow in the <100> directions and the texture effect stems from the shape of the product. HA as well as TDA produce elongated needles or stripe-like structures despite their molar amount in the reaction solution. Compared to the reactions with TAA, the amount of the coligand plays a minor role with TU. As mentioned, the decomposition of TU is more linear over the course of the reaction and therefor at any given time, the same amount of TU is present and the molar abundance between the amine coligand and sulfide stays always the same which leads to similar shapes. Here the same applies as with the TAA findings that the amine favors sulfide rich facets more than lead-rich facets (Table S1).

*Oleic acid as sole ligand*

As mentioned in a previous publication, two-dimensional lead sulfide nanostructures could not be synthesized without chlorine substituted alkane coligands.[8,9,10,11,18,19] Here, we demonstrated that amines and aliphatic fluorides are also suitable for the growth of two-dimensional lead sulfide shapes at reaction temperatures of 170 °C. By omitting the coligand, the only ligand left in the solution is oleic acid. It is therefore surprising that two-dimensional structures were formed with oleic acid as the only ligand (Figure 8). This is probably due to the very high lead to sulfide ratio (140/1) we use in our reactions to favor an oriented attachment to two-dimensional nanostructures instead of a spherical nanoparticle growth. At 170 °C the sulfide source is immediately decomposed and no sulfide is left during the growth step of the nucleated crystals to grow further to spheres. Oriented attachment is then the only possibility to minimize their surface energy. The growth direction of the sheets is <110> and the chloroalkane was responsible to compete for the {110} facets with oleate and in the process destabilizing them, leading to oriented attachment.

At higher temperatures the coligand seems to be less important and the fact that less sulfide monomer is present during the growth step is the most important factor to yield two-dimensional nanosheets. The same reaction performed at 135 °C does not yield nanosheets which supports the claim that the total decomposition or a fast reacting sulfide source is crucial. This is even more supported by our results with TU as the sulfide source where no two-dimensional nanosheets are produced at 170 °C. TU being a more stable sulfide source which decomposes over a longer period of time at this temperature leads to networks of agglomerated star-shaped nanoparticles.

## 4. Conclusion

So far, it was only possible to synthesize lead sulfide nanosheets with a chloroalkane coligand. Here, we demonstrated that other substituted alkanes are also suitable. The chain length of the mentioned alkane plays a minor role but shorter alkanes can produce two-dimensional shapes at lower molar amounts while longer chained alkanes have a smaller molar range in which they yield two-dimensional structures. Most of the obtained nanosheets have a thickness of 10 nm ± 2 nm, which corresponds well with our previous findings where we found that the thickness of the sheets is mostly temperature driven and to a lesser grade controlled by the coligand, when using TAA as the sulfide source.[10] Further, for the first time it was possible to synthesize lead sulfide nanosheets solely with oleic acid and without any other coligand. The reason for this is the higher reaction temperature which leads to a faster decomposition of the sulfide source thioacetamide. Thiourea, as a sulfide source, can also be used to produce nanosheets which are laterally much larger than the nanosheets produced with thioacetamide. Due to the slower decomposition to sulfide ions spherical nanoparticles or other shapes are a significant byproduct.

**Acknowledgments**

The authors thank the German Research Foundation DFG for financial support in the frame of the Cluster of Excellence "Center of ultrafast imaging CUI" and for granting the project KL 1453/9-1. The European Research Council is acknowledged for funding an ERC Starting Grant (Project: 2D-SYNETRA (304980), Seventh Framework Program FP7).

**Figures**

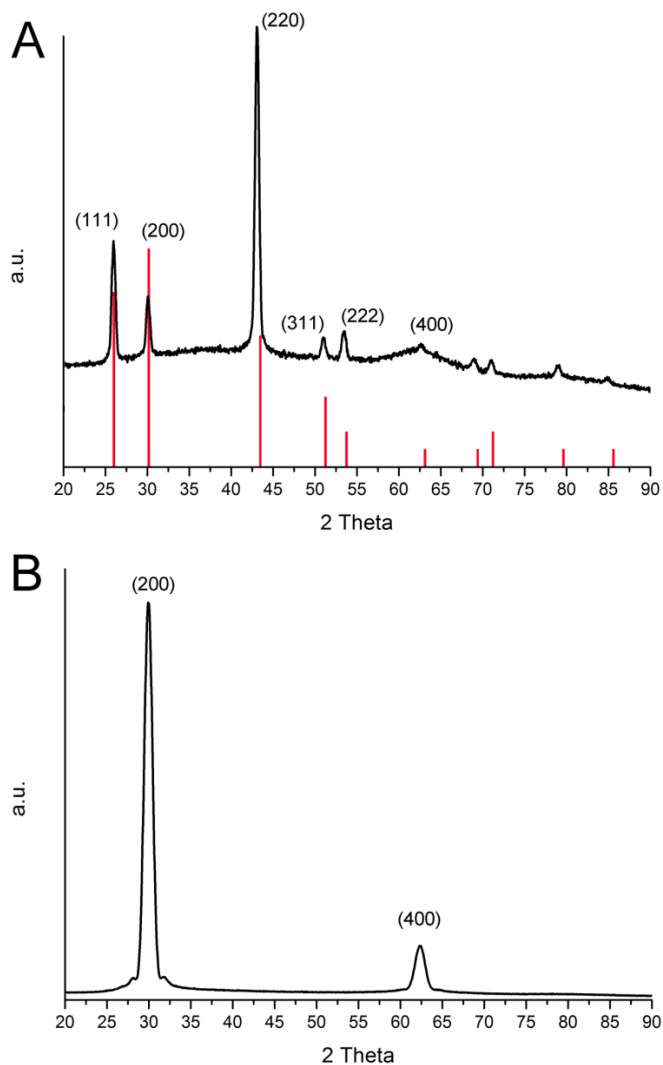

**Figure 1.** *Typical XRD patterns for spherical PbS nanoparticles and nanosheets. (A) shows a typical XRD pattern of syntheses with spherical nanoparticles as the sole product. In this case the reaction was conducted with 0.7 mmol of BH. All galena peaks, (red bars show intensities of a PbS bulk crystal, JCPDS 5-592: galena PbS) are visible in the case of spherical nanoparticles. (B) shows a typical XRD pattern for nanosheets where only the {100} planes are visible due to the texture effect. Such diffractograms were obtained with FH, CH and HA as coligands.*

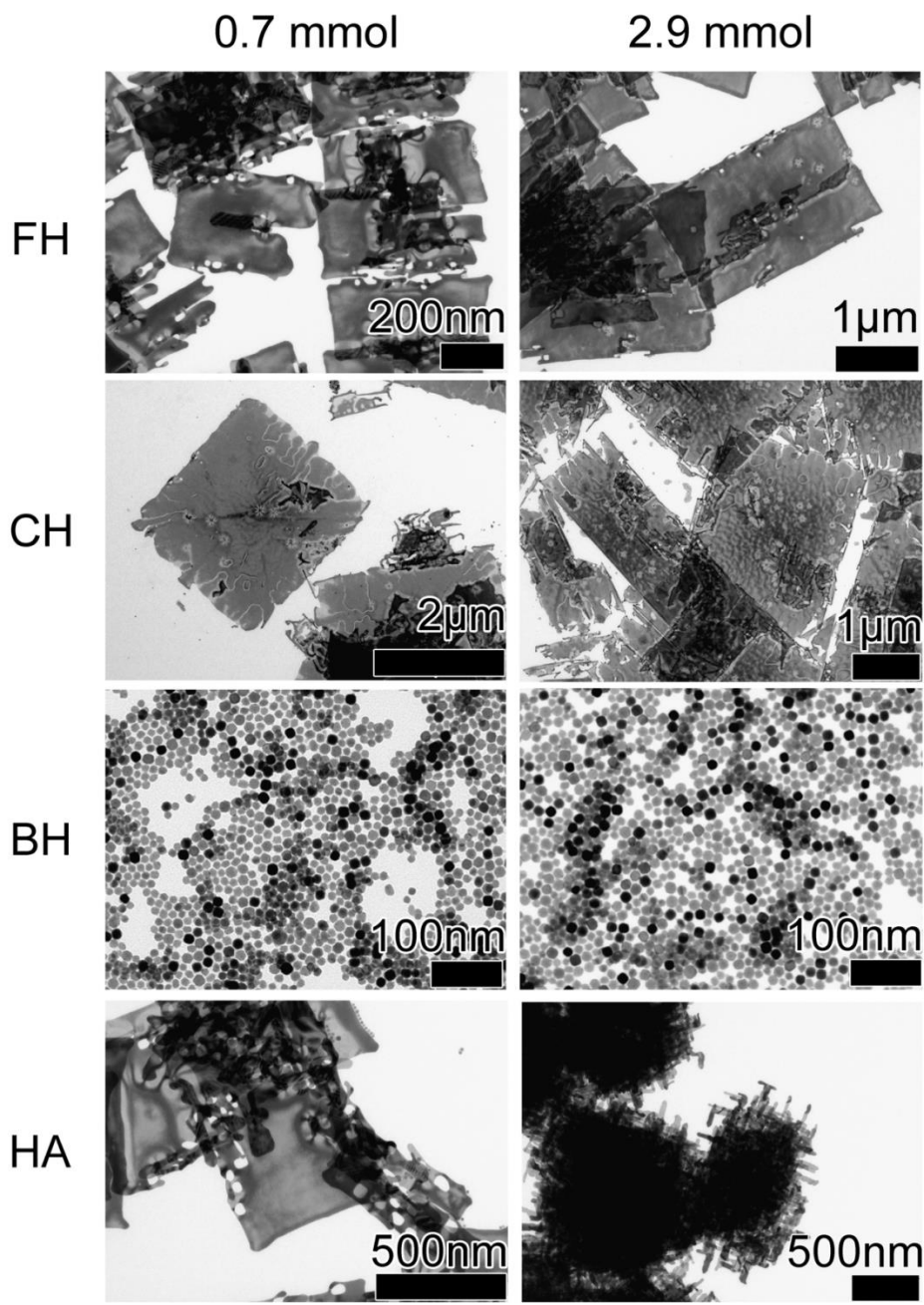

**Figure 2.** *TEM images of the PbS products with different C7 coligands. The reaction temperature was 170 °C for all experiments. Using FH and CH at both molar ratios of 0.7 mmol and 2.9 mmol resulted in nanosheet, while HA yielded nanosheets only at a molar ratio of 0.7 mmol. BH yielded spherical nanoparticles in all cases.*

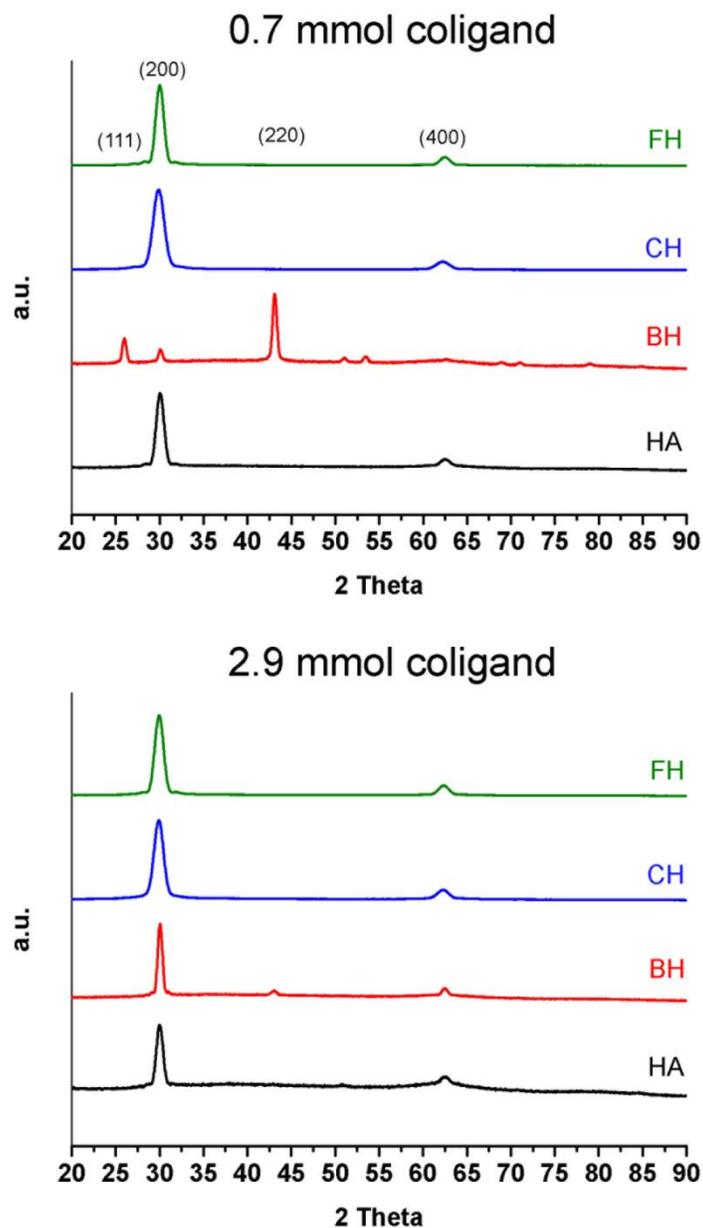

**Figure 3.** *XRD of all products obtained with the shorter C7 ligands. All but the BH XRD show only peaks for the {100} plane of the lead sulfide nanocrystal. This is due to the texture effect of large nanosheets. BH 2.9 mmol could be mistaken for two-dimensional structures but after washing the spherical nanoparticle product several times with toluene the BH XRD changes to bulk galena but the nanosheet XRD stays the same (see Figure S1).*

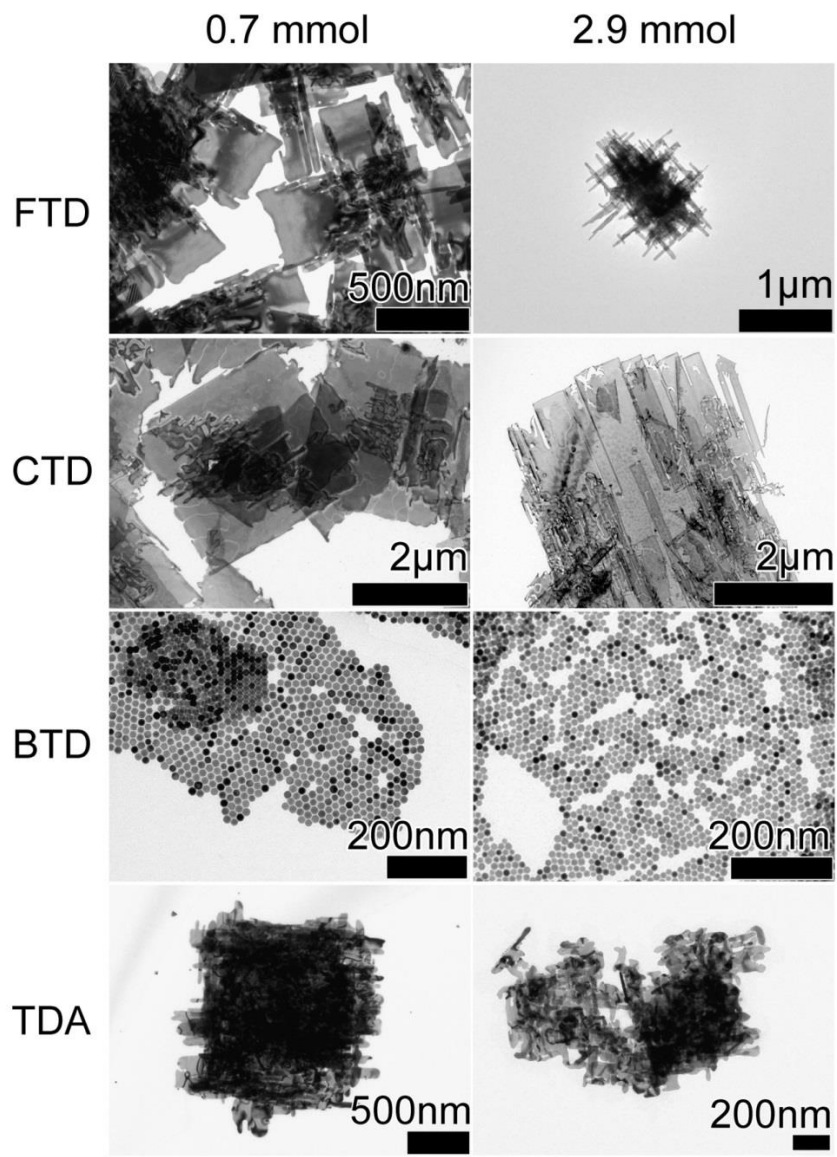

**Figure 4.** *TEM images of reactions with the longer carbon chain (C14) coligands. FTD yields nanosheets only with the 0.7 mmol molar amount and the product becomes agglomerated stripes at 2.9 mmol. With 0.7 mmol of CTD quadratic nanosheets are obtained which become rectangular with one longer lateral dimension with 2.9 mmol CTD. BTD yields spherical nanoparticles just like the shorter BH and TDA yields highly stacked sheets with 0.7 mmol and amorphous platelets with 2.9 mmol.*

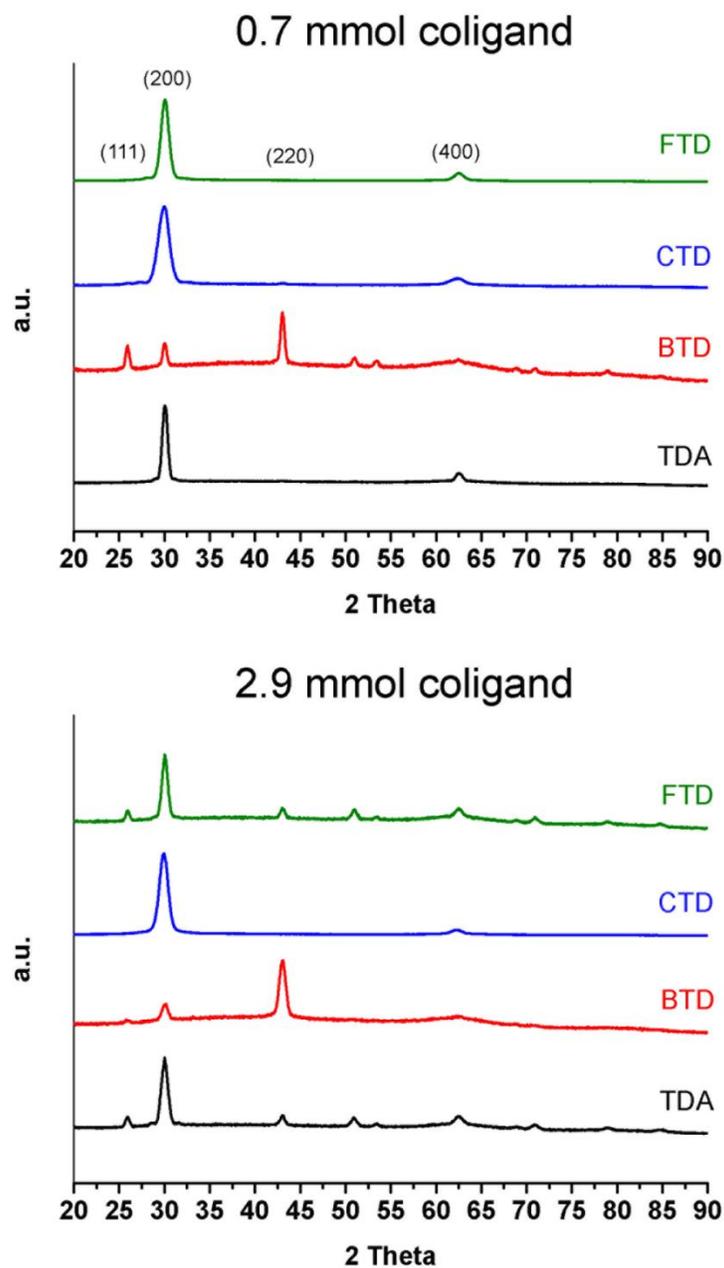

**Figure 5.** *XRD for products obtained with the C14 coligands. While with 0.7 mmol all but BTD show nanosheet characteristic peaks, only CTD shows the same characteristics with 2.9 mmol of the coligand. BTD shows again a ligand driven texture effect which disappears after washing the product several times with toluene (Figure S1).*

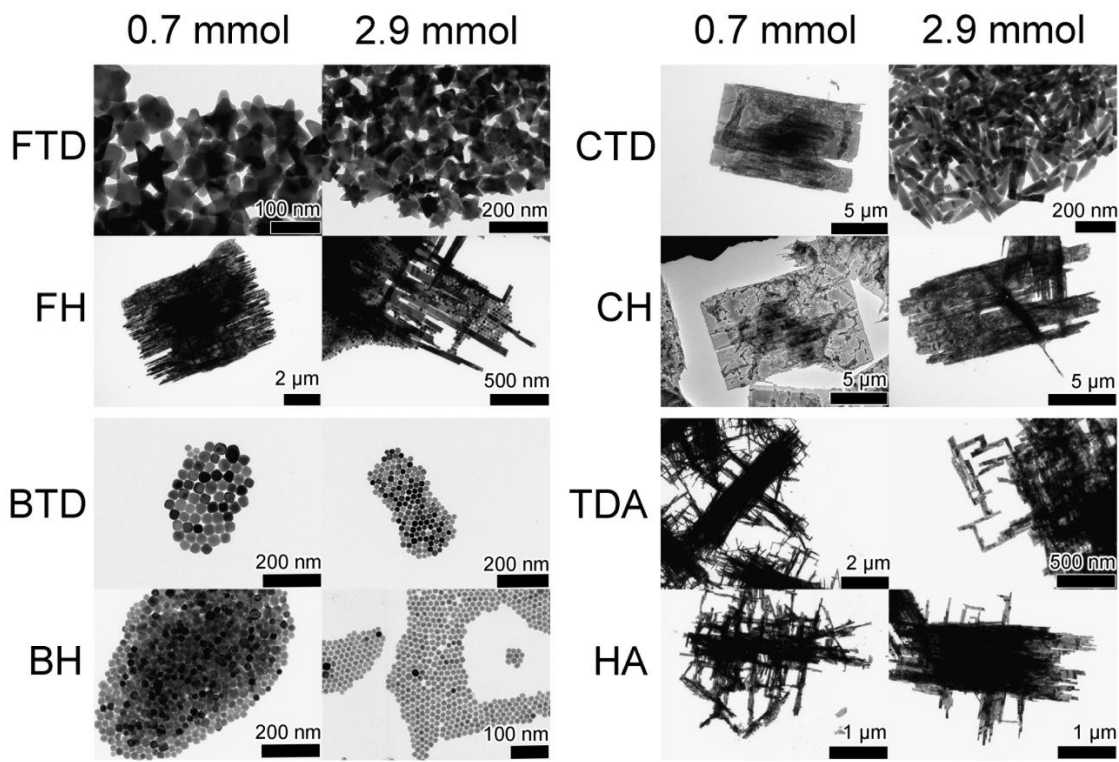

**Figure 6.** *TEM images of reactions where TU was used instead of TAA as a sulfide source to yield PbS nanocrystals. FTD only yielded star-shaped structures. The shorter fluorine ligand (FH) yielded nanosheets at the lower concentration which are very large in diameter (6x6 µm²) but these sheets consist of grown-together stripes and with the higher concentration the stripe-like structures are more common in the product. The chloro coligands CTD and CH yielded both large nanosheets (7x10 µm²). With 2.9 mmol of CTD only bolt-like structures are obtained and with 2.9 mmol CH the nanosheets are more stripe-like and elongated, bolt-like structures are obtained as a byproduct, too. The bromoalkanes only yield spherical nanoparticles regardless of the amount used. The amines yield very stripey nanosheets where with the shorter HA the nanosheet structure is more apparent than with the TDA.*

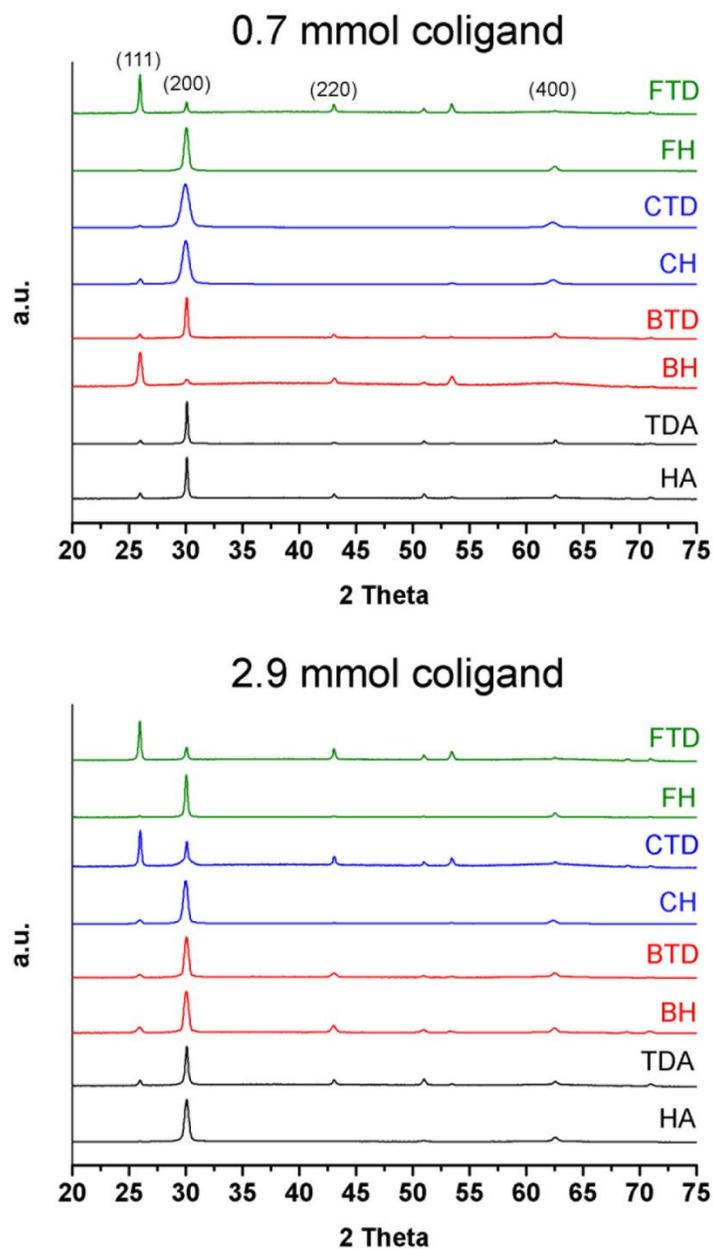

**Figure 7.** *Shown are all XRD for the reaction with TU instead of TAA as the sulfide source. Despite TEM images showing spherical nanoparticles, nanosheet only peaks outbalance the signals for spherical nanoparticles by far. This is because the produced nanosheets with FH and CH, for both molar amounts, as well as for CTD at 0.7 mmol, are much larger than the TAA sheets and induce a higher intensity for the {100} factes.*

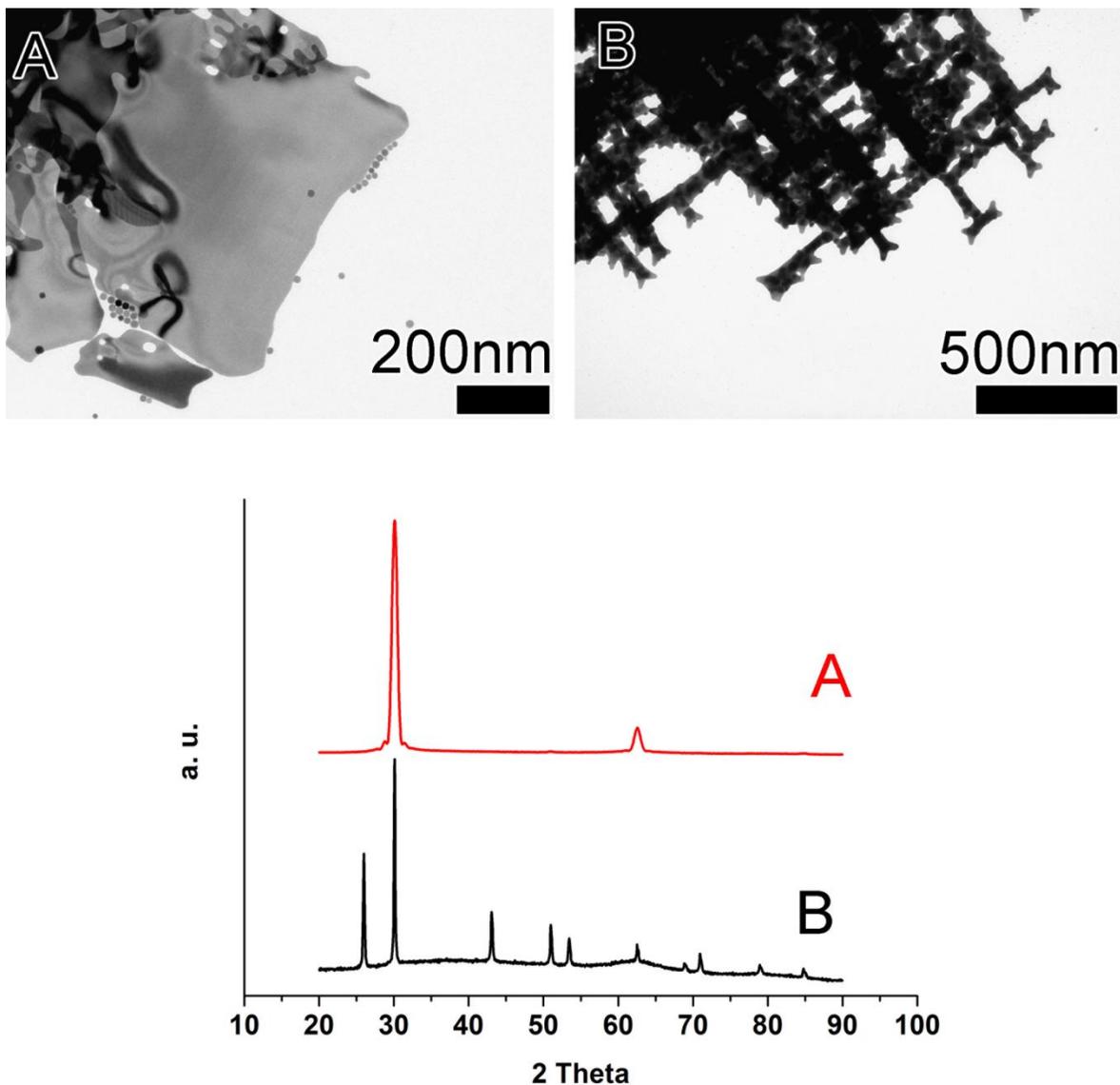

**Figure 8.** *TEM images of lead sulfide nanosheets obtained without any coligand and the corresponding XRD. The TEM image (A) shows two-dimensional structures which were obtained with oleic acid as the sole ligand and TAA as the sulfide source at a reaction temperature of 170 °C. XRD also supports this finding with peaks showing {100} planes only. (B) on the other hand shows agglomerated networks of star-shaped structures which exhibit all galena peaks in the XRD. The difference here was that the used sulfide source was TU and not TAA.*

Supporting Information

**New Ways to Synthesize Lead Sulfide Nanosheets –**

**Substituted Alkanes Direct the Growth of 2D Nanostructures**

Thomas Bielewicz, Eugen Klein, and Christian Klinke*

*Institute of Physical Chemistry, University of Hamburg,*

*Grindelallee 117, 20146 Hamburg, Germany*

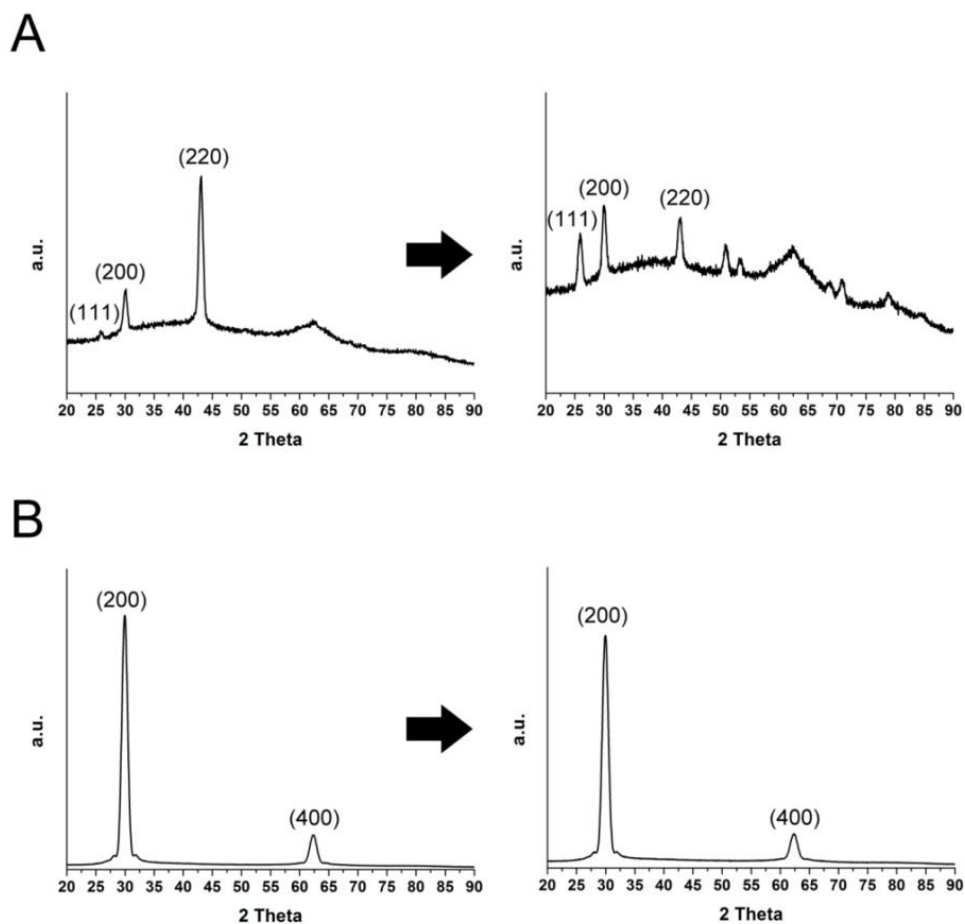

*Figure S1: XRD patterns before and after washing PbS spherical nanoparticles (A) or sheets (B). (A) shows a XRD pattern of spherical nanoparticles synthesized with bromoheptane (2.9 mmol). After washing a part of the product again two times with toluene the XRD pattern changes to a bulk galena distribution. The texture effect was therefore not caused by the shape but by excessive ligands which were then washed away. (B) on the other hand shows no change in the XRD after washing nanosheets two more times with toluene. Here the texture effect is caused by the shape of the product and washing ligands away does not change the alignment of the sheets on the substrate by dropcasting.*

**Table S1: Adsorption energies of different coligand and ligand molecules for the most important PbS crystal facets calculated by the DFT method.** The simulations were performed using shorter versions of the coligand and ligand molecules with the functional group and a three carbon long chain.

| adsorption energy (eV) | PbS-110 | PbS-100 | PbS-111 (lead rich) | PbS-111 (sulfur rich) |
|---|---|---|---|---|
| propanoate | 3.98 | 3.32 | 4.47 | 1.96 |
| propylamine | 0.97 | 0.88 | 0.56 | 0.90 |
| fluoropropane | 0.80 | 0.49 | 0.34 | 0.40 |
| chloropropane | 0.91 | 0.56 | 0.40 | 0.47 |
| bromopropane | 0.97 | 0.63 | 0.43 | 0.52 |
| iodopropane | 1.07 | 0.73 | 0.48 | 0.66 |

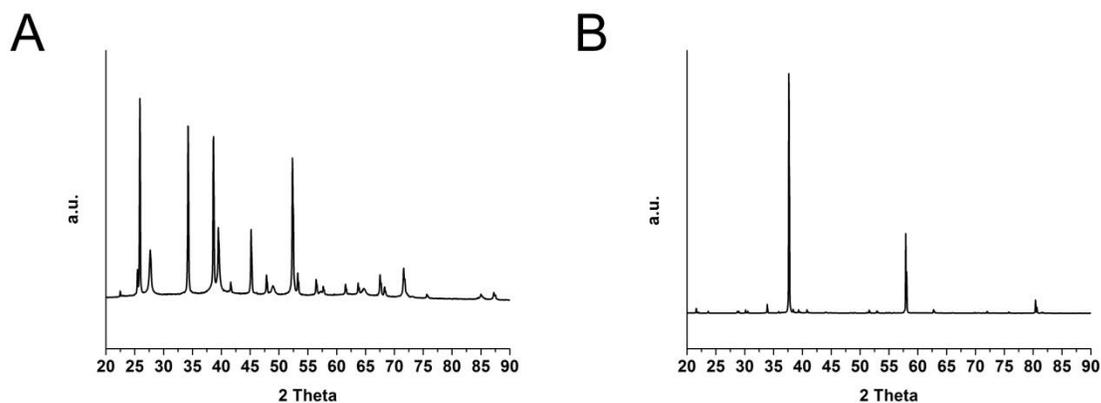

***Figure S2: XRD of PbI₂ and PbBr₂.*** *(A) shows the XRD of the yellow PbI$_2$ obtained in the reaction with 2.9 mmol of iododecane and (B) shows the XRD of the white PbBr$_2$ obtained with bromotetradecane as the solvent (10 mL). In both cases TAA was also in the reaction solution but as shown here no galena formed.*

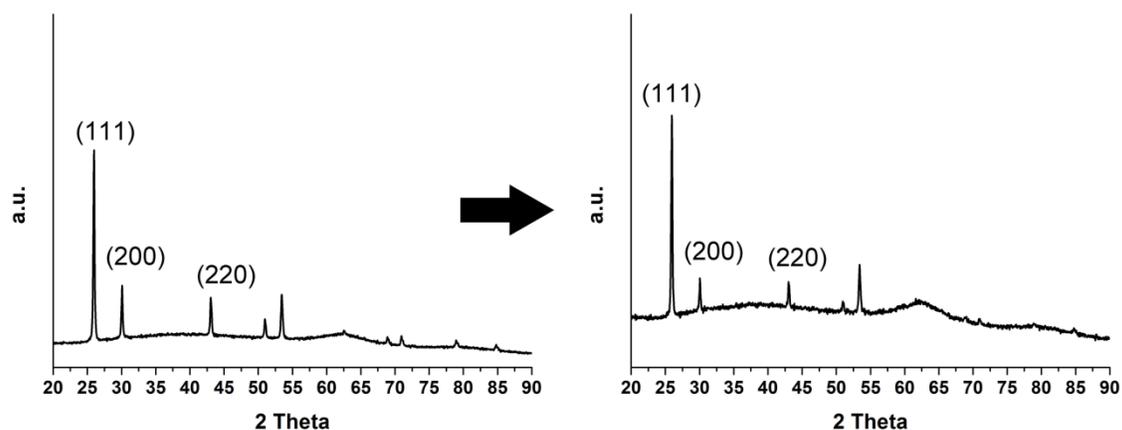

***Figure S3: XRD for star shaped product obtained with FTD and thiourea.*** *The left XRD shows the star shaped nanostructures after washing two times with toluene. The right XRD shows the product after washing two more times with toluene and removing excess ligands in the process. The texture effect still remains with the (111) plane being the most prevalent. Here, as well as with sheets, the shape is inducing the texture effect.*